\documentclass[11pt,onecolumn]{IEEEtran}

\input{macros}

\headheight     = \baselineskip
\headsep        = \baselineskip
\textheight     = \paperheight
\textwidth      = \paperwidth
\linewidth      = \textwidth

\title{Semantic Guidance and Feedback for the Construction of
       Specifications and Implementations}

\author{\IEEEauthorblockN{Paul C. Attie and Fadi A. Zaraket and Mohammad Fawaz and Mohamad Noureddine} \\
\IEEEauthorblockA{American University of Beirut \\
    Email: \{pa07,fz11,mbf12,man17\}@aub.edu.lb}
}

\begin{document}
\maketitle



\begin{abstract}
The problem of writing a specification which accurately reflects the intent of the developer has
long been recognized as fundamental.  We propose a method and a supporting tool to write and check a
specification and an implementation using a set of use cases, \ie input-output pairs that the
developer supplies.  These are instances of both good (correct) and bad (incorrect) behavior. We
assume that the use cases are accurate, as it is easier to generate use cases than to write an
accurate specification.  We incrementally construct a specification (precondition and postcondition)
based on semantic feedback generated from these use cases.  We check the accuracy of the constructed
specification using two proposed algorithms. The first algorithm checks the accuracy of the
specification against an automatically generated specification from a supplied finite domain of use
cases.  The second checks the accuracy of the specification via reducing its domain to a finite yet
equally satisfiable domain if possible.  When the specification is mature, we start to also
construct a program that satisfies the specification. However, our method makes provision for the
continued modification of the specification, if needed.  We illustrate our method with two examples;
linear search and text justify.
\end{abstract}

\section{Introduction}

The derivation of programs from formal specifications, and the construction of a correctness proof
hand-in-hand with the program has been advocated by Dijkstra \cite{Dij76}, Hoare \cite{Ho69}, Gries
\cite{Gr81}, and others.  Central to this method is the a priory existence of a formal
specification, that is assumed to be accurate, \ie representing what the user requires.  The task of
constructing such a specification is addressed by the many requirements elicitation methods that
have been presented in the
literature~\cite{HWT03,HST98,HKLAB98,KKK95,CABetal98,F89,HJL96,MH91,GGH90,MS03}.  See
Section~\ref{sec:related} below for a discussion of some of these.

\paragraph{Our contributions}
We present a manual method for the construction of specifications for data-intensive,
transformational, terminating sequential programs.  Our method deals with sequential imperative
programs, written in any standard sequential OO language, \eg Java or C++.  Currently, we support
arithmetic and Boolean expressions, scalar and array data types, assignment, if, while, and
procedure calls.  It is a simple matter to extend our implementation to other constructs, see
Section~\ref{s:tool}.  Specifications consist of a precondition, postcondition pair, written in
first order logic.  There is a single input, which is restricted by the precondition, and a single
output, which is related to the input by the postcondition.
Given a set of use cases, the method evaluates the specification
and returns {\em correction actions} which the user executes to 
correct the specification. 

We also present two algorithms for checking the resulting specification.  The first checks the
accuracy of the specification via reducing its domain to a finite domain which preserves
satisfaction.  The second checks the accuracy of the specification against an automatically
generated specification from a supplied finite domain of the input and output variables.  In many
cases, this will reveal deficiencies in the manually written specification.

We have implemented our method, available online~\footnote{\label{fn:online}
  \url{http://webfea.fea.aub.edu.lb/fadi/dkwk/doku.php?id=speccheck}}, and all the examples given
below were constructed using our implementation.

The rest of the paper is organized as follows.  We present the method in Section~\ref{s:method}, and
the linear search example in Section~\ref{s:ex}.  
Section~\ref{s:overimpl} discusses additional implementation correction actions.  We
describe the supporting tool in Section~\ref{s:tool}, provide the text justify example in
Section~\ref{s:justify}, and discuss related work in Section~\ref{s:related}.  We conclude and
discuss future work in Section~\ref{s:conclusion}.

\section{The Method}
\label{s:method}


\subsection{Behaviors and specifications}

\begin{table*}[bt]
\centering
\tiny
\caption{ Adequacy checks for precondition $P$. }
\vspace{1ex}
\resizebox{1.0\columnwidth}{!}{
\begin{tabular}{c|c|c|c|p{2.5cm}} 
                 & $P(i)$     & Actual       & Required       & Correction Action\\ \hline
    \multirow{2}{*}{\tgood $\un$ \tbad}
                & \true     & $P$            & $P$            & skip \\ \cline{2-5}
                & \false    & $\neg P$       & $P$            & $P \up i$ \\ \hline \hline

    \multirow{2}{*}{\tdc}
                & \true     & $P$            & $\neg P$       & $P \dn i$ \\ \cline{2-5}
                & \false    & $\neg P$       & $\neg P$       & skip \\  \hline
\end{tabular}
}
\normalsize
\label{t:pchecks}
\end{table*}

\begin{table*}[bt]
\centering
\tiny
\caption{ Adequacy checks for postcondition $Q$. }
\vspace{1ex}
\resizebox{1.0\columnwidth}{!}{
\begin{tabular}{c|c|c|c|p{2.5cm}} 
                & $Q(i,o)$  & Actual         & Required       & Correction Action\\ \hline
    \multirow{2}{*}{\tgood}
                & \true     & $Q$            & $Q$            & skip \\ \cline{2-5}
                & \false    & $\neg Q$       & $Q$            & $Q \up (i,o)$ \\ \hline \hline

    \multirow{2}{*}{\tbad}
                & \true     & $Q$            & $\neg Q$       & $Q \dn (i,o)$ \\ \cline{2-5}
                & \false    & $\neg Q$       & $\neg Q$       & skip \\  \hline
\end{tabular}
}
\normalsize
\label{t:qchecks}
\end{table*}

The input of a program is given by a fixed set of \emph{input variables} 
$\bar{x} = \tpl{x_1,\ldots,x_m}$, \eg actual
parameters. Likewise, the output of a program is given by a fixed set of \emph{output
variables} $\bar{y} = \tpl{y_1,\ldots,y_n}$, \eg parameters that are reference types, or a single variable whose value is
passed to a return statement. A variable can be both input and output, so the $\bar{x}$ and
$\bar{y}$ are really meta-syntactic variables. Each $x_j$, $j = 1, \ldots, m$, takes
values from a domain $I_j$, and each $y_j$, $j = 1, \ldots, m$,  takes
values from a domain $O_j$.
Then $I = I_1 \times \cdots \times I_m$ is the
domain of the inputs, and $O = O_{1} \times \cdots \times O_n$ is the
domain of the outputs.

A \emph{behavior} is a pair $(i,o)$ where $i \in I, o \in O$. 
$i = \tpl{iv_1,\ldots,iv_m}$ is a tuple of the initial values of the input variables, and 
$o = \tpl{ov_{1},\ldots,ov_n}$ is a tuple of the final values of the output.
We partition the set $I \times O$ of behaviors into three sets:
\be
\i $\good$, the set of good (positive) behaviors: the precondition holds before and the
postcondition holds after
\i $\bad$,  the set of bad (negative) behaviors: the precondition holds before and the postcondition
does not hold after
\i $\dc$, the set of don't care behaviors: the precondition does not hold before, and the
postcondition can be either true or false after
\ee

We assume that the developer can reliably provide use cases and classify them
as good, bad, and dontCare behaviors.

The set $I \times O$ of behaviors is, in general, infinite, and so cannot be represented
directly. The usual representation is a \emph{specification} $\Sp$, consisting of a precondition
$P(i)$ over the input and a postcondition $Q(i,o)$ over the input and output.  We write
preconditions and postconditions in first order logic. We will assume standard terminology, such as
subformula, and atomic formula (quantification or a relation applied to terms), and free variable. 
Let $f$ be a wff of first order logic. Then we write 
$vars(f)$ for the set of free variables that occur in $f$. 
We write $f' < f$ iff $f'$ is a subformula of $f$.
We write $at(f)$ for the set of wff's $f'$ such that (1) $f'$ is a
subformula of $f$, (2) $f'$ is atomic (\ie a quantification or a relation symbol applied
to terms) and (3) there does not exist an atomic wff $f''$ such that $f' < f'' < f$. In words, $f'$
is a ``maximal'' atomic subformula of $f$.
We extend these notions to specifications:
$vars(\Sp) \df vars(P) \un vars(Q)$, and $at(\Sp) = at(P) \un at(Q)$.
We assume, without loss of generality, that all the input and output variables 
occur free in $\Sp$, \ie $\bar{x} \un \bar{y} \sub vars(\Sp)$.

We will, as convenient, write the truth value of forumla $f$ within a valuation (model) $V$ as
$V(f)$ or as $f(V)$, 
ie sometimes we view valuations as mapping formulae to \tf, and sometimes we view formulae as
mapping valuations to \tf. To reduce use of parenthesis, we sometimes use dot notation: $V.f$, $f.V$,
respectively. 

It is recognized that writing specifications is difficult because it involves formalizing
an informal idea of behavior, given as a set of behaviors (use cases). 
Our aim in this paper is to bring formality and rigour to the task of writing specifications.

\subsection{Definitions}

\begin{definition}
\label{def:spec-general}
A behavior $(i,o)$ \emph{satisfies} a specification $\Sp = (P,Q)$ iff $P(i) \imp Q(i,o)$. We
write $(i,o) \sat \Sp$. We also write $[\Sp] \df  \{ (i,o) : (i,o) \sat \Sp \}$.
A specification $\Sp$ is \emph{under-constrained} iff 
there exists $(i,o) \in \bad$ such that $(i,o) \sat \Sp$.
A specification $\Sp$ is \emph{over-constrained} iff 
there exists $(i,o) \in \good$ such that $(i,o) \not\sat \Sp$.
A specification is \emph{accurate} iff it is not under-constrained and not over-constrained.
\end{definition}

It is immediate that, for an accurate specification $\Sp$:
$\good \sub [\Sp]$ and $\bad \ints [\Sp] = \emptyset$.
There is no restriction on the relation between $[\Sp]$ and $\dc$. Hence, given 
a partition of the behavior space $I \times O$ into $\good$, $\bad$, and $\dc$, there are many
possible specifications that correspond to this partition, depending on how much of the $\dc$
partition is included in $[\Sp]$.


Crucial to our method is the ability to evaluate $P(i)$ and $Q(i,o)$
given $(i,o)$. This presents no problem, provided that
all quantifications have finite (but unbounded, \eg the size of an input array) range. 
We assume this restriction in the sequel.

\begin{definition}
Let $R, R'$ be predicates.
$R'$ is {\em stronger} than $R$ iff $R' \imp R$ is valid. 
$R'$ is {\em weaker} than  $R$ iff $R \imp R'$ is valid.
\end{definition}

\subsection{Correcting a specification using behaviors}

We generate, by hand, several behaviors $(i,o)$, 
   both \good and \bad.
We evaluate $P(i)$ and $Q(i,o)$ for every pair $(i,o)$. 
The results generate \emph{correction actions}, as given in 
Tables~\ref{t:pchecks} and ~\ref{t:qchecks}.
A correction action is a Boolean formula, where the propositions are
\emph{basic correction actions}, as follows.
$R \dn v$ means strengthen predicate $R$ so that $R(v) = \false$. 
$R \up v$ means weaken $R$ so that $R(v) = \true$. 
$skip$ means leave $\Sp$ unchanged.

Table~\ref{t:pchecks} deals with the precondition $P$. The ``Actual'' column gives the outcomes
from evaluating $P$, and the ``Required'' column gives the expected outcome. For 
good pairs, $P$ must hold and for bad pairs, $\lnot P$ must hold. 
The ``Correction Action'' column gives the correction action that can be applied to $P$.
The top half of the table shows the correction action for both good and bad behaviors, while
the lower half shows the correction actions for the dontCares.

Table~\ref{t:qchecks} shows similar information for the postcondition $Q$. The top half of the table
deals with good behaviors while the lower half deals with bad behaviors. Note that we do not need 
correction actions in case of a dontCare behavior since by definition, the postcondition in a 
dontCare behavior can either be \true~or \false. 

Consider the first row in Table~\ref{t:pchecks}. 
The precondition $P$ actually holds, which is also the required value. Therefore
no correction action is needed and the user is instructed to skip, i.e. leave $P$ as it is. The second row 
shows the case where $P$ does not actually hold while it is required to do so. In this case, the correction
action requires the user to weaken the precondition $P \up i$ so that it holds over the input $i$. The rest
of this table as well as Table~\ref{t:qchecks} is similarly explained. 



%
%

\subsection{Correcting both a specification and an implementation using behaviors}


We now add an implementation $S$. We assume that $S$ is sufficiently developed that it can
generate outputs for any input.
Define $g(i,o) \mbox{ iff } (i,o) \in \cci{good}$, 
and $S(i,o)$ iff $o$ is a result of execution $S$ with input $i$.
Also define basic corrective action
$S \dn (i,o)$: modify $S$ so that execution of 
$S$ with input $i$
does not produce output $o$.

Since $S$ generates output, 
      we use inputs $i$ instead of behaviors $(i,o)$.
Given input $i$ such that $P(i)$, we execute $S$ with input $i$, resulting in some output $o$. 
Table~\ref{t:psqchecks} gives the appropriate correction action for each case.
The column labeled $g(i,o)$ gives the value of $g(i,o)$, which is input by an interactive query to the 
developer, who is the ultimate reference for \good and \bad.
The column labeled  $Q(i,o)$ gives the value 
obtained by evaluating $Q(i,o)$.
For example, the third row has $\neg g(i,o)$ and $Q(i,o)$, 
    \ie $o$ is not a
good output according to the developer, but it does satisfy the postcondition. 
The correction action $Q \dn (i,o) \land S \dn (i,o)$
thus requires that we change $S$ so that $o$ is not produced on input
$i$, and we also change $Q$ so that $Q(i,o)$ is false, \ie $Q$ reflects that
$o$ is an incorrect output for input $i$.

\begin{table}[bt]
\centering
\tiny
\caption{ Adequacy of $\{P\} S \{Q\}\{B\}$ for $P(i)$. }
\vspace{1ex}
\resizebox{1.0\columnwidth}{!}{
\begin{tabular}{c|c|p{3.0cm}} 
  $g(i,o)$    & $Q(i,o)$  & Correction Action\\ \hline
\multirow{2}{*}{\true}
                 & \true      & skip \\ \cline{2-3}
                 & \false     & $Q \up (i,o)$ \\ \hline
\multirow{2}{*}{\false}
                 & \true       & $Q \dn (i,o) \land S \dn (i,o)$ \\ \cline{2-3}
                 & \false       & $S \dn (i,o)$ \\  \hline 
\end{tabular}
}
\normalsize
\label{t:psqchecks}
\end{table}

\section{The \cci{linearSearch} example}
\label{s:ex}
We illustrate the method using a linear search example.
Function \cci{linearSearch} takes as input an array \chci{a},
indices \chci{l} and \chci{r} that define left and right boundaries of the search, respectively,
and element \chci{e}. 
\cci{linearSearch} returns the
index of \chci{e} in \chci{a} if \chci{e} was found between \chci{l} and \chci{r} inclusive, and returns $-1$ otherwise.
The following code listing shows the interface (or the empty definition) of \cci{linearSearch}. \\
\begin{tabular}{p{8cm}}
 \begin{Verbatim}[fontsize=\relsize{-2},numbersep=4pt,numbers=left,frame=topline,framesep=4mm,label=\fbox{ Linear search interface}]
int linearSearch(int [] a, int l, int r, int e);
\end{Verbatim} 
\\
\end{tabular} 
\\
We start with the weakest specification, $(\false, \true)$ (\ie $P =
\false$ and $Q = \true$), which
admits all behaviors. We repeatedly refine this specification by
considering a single behavior, computing the actual values of 
$P$ and then $Q$ on the behavior, comparing with the expected value
(depending on whether the behavior is good or bad), and then applying
the corresponding correction action from Tables~\ref{t:pchecks} and~\ref{t:qchecks}
for $P$ and $Q$ respectively.



\noindent
\empb{Pair 1} (good, required $P \land Q$): \\
input:   \cd{a=\{1,2,3\}, l=0, r=2, e=4}\\
output:  \cd{rv=-1}\\
actual: $\neg P \land Q$\\
correction: ($P \up $i) for $P$ and skip for $Q$ \\
User intuition dictates that $Q(i,o)$ should hold, since the output
\cd{rv=-1} correctly indicates that the value \cd{4} is not present in \cd{a}.
From Table~\ref{t:pchecks}, the correction action for $P$ is to weaken it 
so that it holds for the given input. Array \cd{a} and search element 
\cd{e} can have arbitrary values, so we cannot constrain them in the 
precondition. We note that it makes sense for \cd{l} to be $\le$ \cd{r},
since otherwise the search interval is empty. Thus we weaken $P$ to 
$l \le r$. $Q$ remains $\true$ since the correction action from Table~\ref{t:qchecks}
is to skip. 

\noindent
\empb{Pair 2} (good, required $P \land Q$): \\
input:   \cd{a=\{1,2,3,4,5\}, l=0, r=4, e=2}\\
output:  \cd{rv=1}\\
actual: $P \land Q$\\
correction: skip for both $P$ and $Q$\\
No changes are made for this $(i,o)$ pair as dictated by Tables~\ref{t:pchecks} and~\ref{t:qchecks}.

\noindent
\empb{Pair 3} (bad, required $P \land \neg Q$): \\
input:   \cd{a=\{1,2,3\}, l=0, r=2, e=4}\\
output:  \cd{rv=0}\\
actual: $P \land Q$\\
correction: skip for $P$ and $Q\dn(i,o)$ for $Q$ \\ 
Now, we must strengthen $Q$ so that $Q(i,o)$ does not hold.
Our informal description of the linear search problem states that when
\cd{rv} is not -1, then it must give a location of \cd{e} in
\cd{a}. Hence we strengthen $Q$ to: $rv \ne -1 \imp a[rv] = e$.
$P$ remains $l \le r$.\\

\noindent
\parbox{\columnwidth}{
\empb{Pair 4} (bad, required $P \land \neg Q$): \\
input:   \cd{a=\{5,2,7,3,6,8\}, l=1, r=4, e=7}\\
output:  \cd{rv=-1}\\
actual: $P \land Q$\\
correction: skip for $P$ and $Q\dn(i,o)$ for $Q$ \\ 
Now, we must strengthen $Q$ so that $Q(i,o)$ does not hold.
Our informal description of the linear search problem states that when
\cd{rv} is -1, then \cd{e} does not occur in
\cd{a}. Hence we strengthen $Q$ to: 
$(rv \ne -1 \imp a[rv] = e) \land (rv = -1 \imp (\fa k : 0 \le k < a.size \imp e \ne a[k]))$.
$P$ remains $l \le r$.\\
}

\noindent
\empb{Pair 5} (dontCare, required $\neg P$): \\
input:   \cd{a=\{5,2,7,3,6,8\}, l=4, r=1, e=7}\\
output:  \cd{rv=-1}\\
actual: $\neg P \land \neg Q$\\
correction: skip for $P$, no action (i.e. skip) for $Q$ \\ 
User intuition dictates that $P(i)$ should remain $\false$, since the
search interval is empty, which is dictated by Table~\ref{t:pchecks}.
No action should be taken on the postcondition $Q$.

\noindent
\empb{Pair 6} (good, required $P \land Q$): \\
input:   \cd{a=\{5,2,7,3,6,8\}, l=0, r=1, e=7}\\
output:  \cd{rv=-1}\\
actual: $P \land \neg Q$\\
correction: skip for $P$ and $Q \up (i,o)$ for $Q$ \\
The user must now weaken the postcondition $Q$ so that $Q(i,o)$ holds.
The conjunct $(rv \ne -1 \imp a[rv] = e)$ of $Q$ holds vacuously.
The conjunct $(rv = -1 \imp (\fa k : 0 \le k < a.size \imp e \ne a[k]))$.
fails to hold since \cd{e} occurs
in \cd{a}, at position 2. However, the search interval is $0$ to $1$,
and the occurrence of \cd{e} is outside of the interval. Hence, we
should incorporate the search interval into $Q$. We might as well do
this for both conjuncts now, even though this particular case does not
require correction of the conjunct $(rv \ne -1 \imp a[rv] = e)$,
other cases will. The revised $Q$ is 
$(rv \ne -1 \imp l \le rv \le r \land a[rv] = e) \land (rv = -1 \imp (\fa k : l \le k \le r \imp e \ne a[k]))$.
$P$ is unchanged.\\

\noindent
\empb{Pair 7} (good, required $P \land Q$): \\
input:   \cd{a=\{5,2,7,3,6,8\}, l=-1, r=10, e=7}\\
output:  \cd{rv=-1}\\
actual: $P$ holds, $Q$ is undefined.\\
This brings up an issue that is not addressed by Tables~\ref{t:pchecks} and~\ref{t:qchecks}: 
what if $P$ or
$Q$, as currently formulated, cannot be evaluated for some behavior? In this case, the
problem is that the search interval $l, \ldots, r$ extends outside the index range of
array \cd{a}. For such cases, we define the appropriate correction to be that which
results in $P$ and $Q$ being well-defined. Hence, we strengthen $P$ to 
$0 \le l \le r < a.size$, and we strengthen $Q$ to
$0 \le l\le r < a.size \land 
(rv \ne -1 \imp l \le rv \le r \land a[rv] = e) \land
(rv = -1 \imp (\fa k : l \le k \le r \imp e \ne a[k]))$.

Since the specification is mature, we add our 
implementation of \cci{LinearSearch} annotated with the 
specification and the input-output pairs as shown in the following
code listing. 

\begin{tabular}{p{8cm}}
  \begin{Verbatim}[fontsize=\relsize{-2},numbersep=4pt,numbers=left,
    frame=topline,framesep=4mm,
    label=\fbox{ Linear search annotated},
commandchars=\\\{\}, codes={\catcode`$=3\catcode`_=8}]
int LinearSearch(int [] a, int l, int r, int e) \{
   @pre ls (0 $\leq$ l $\leq$ r < a.size);
    int i = l;
    while ( i $\leq$ r ) \{
        if (a[i] == e)
            break;
        i++; \}
    return -1;
   @post ls \{
      (0 $\leq$ l $\leq$ r < a.size)
      ( (rv $\not=$ -1) $\imp$ l $\leq$ rv $\leq$ r $\land$ a[rv] = e)
      ( (rv = -1) $\imp$ $\forall$ int k:[l .. r] (e $\not=$ a[k])) \}

   @behavior ls \{
      good \{ input=\{a=\{1,2,3\},       l=0, r=2, e=4\} 
             output=\{rv=-1\} \}
      good \{ input=\{a=\{1,2,3,4,5\},   l=0, r=4, e=2\} 
             output=\{rv=1\} \} $\ldots$ \} \}
\end{Verbatim}
\\
\end{tabular}

We proceed to the $\{P\} S \{Q\}\{B\}$ checks of 
Table~\ref{t:psqchecks}.
The input from Pair 2 invokes the correction $S \dn (i,o)$
from the fourth row of Table~\ref{t:psqchecks}. 
Indeed our implementation computes the correct result but returns 
-1 instead in all cases. 
We correct the implementation and replace the \cci{break;}
with a \cci{return i;} statement. 

We now consider the use of \cci{LinearSearch} as a priority queue
where the key and the position of the elements have semantic
significance and propose our expectation that \cci{LinearSearch}
returns the {\em rightmost match} of \chci{e} if it exists using
the following two input-output pairs. 

\begin{tabular}{p{8cm}}
 \begin{Verbatim}[fontsize=\relsize{-2},numbersep=4pt,numbers=left,firstnumber=16,
commandchars=\\\{\}, codes={\catcode`$=3\catcode`_=8}]
   $\ldots$
      good \{ input=\{a=\{5,2,7,6,7,8\}, l=1, r=5, e=7\}
            output=\{rv=4\} \}
      bad  \{ input=\{a=\{5,2,7,6,7,8\}, l=1, r=5, e=7\}
            output=\{rv=2\} 
\end{Verbatim}
\\
\end{tabular}

The program computes $2$ as the index and the postcondition
$Q(i,o)$
evaluates to $\true$.
However, $\neg g(i,2)$ and Table~\ref{t:psqchecks} recommends 
the third row $Q \dn (i,o) \land S \dn (i,o)$ 
requiring that we strengthen $Q$ to refute $2$ and 
strengthen the implementation to return $4$. 
We strengthen the second conjunct of $Q$ to 
$(rv \ne -1 \imp ( 
        (l \le rv \le r  \land a[rv] = e) \land 
        (\forall k. (k > rv \land k \le r) \imp a[k] \not= e)
        )$.
We also strengthen the implementation to search \chci{a}
from right to left as follows.

\begin{tabular}{p{8cm}}
 \begin{Verbatim}[fontsize=\relsize{-2},numbersep=4pt,numbers=left,firstnumber=4,
    frame=topline,framesep=4mm,
    label=\fbox{ Linear search, return rightmost match},
commandchars=\\\{\}, codes={\catcode`$=3\catcode`_=8}]
    int i = r;
    while ( i $\geq$ l ) \{
        if (a[i] == e)
            return i;
        i--; \}
    return -1;
\end{Verbatim}
\\
\end{tabular}



\vspace{-.4em}
\section{The over-implementation check }
\label{s:overimpl}

\begin{table}[bt]
\centering
\tiny
\caption{ Adequacy of $\{P\} S \{Q\}\{B\}$ for $\neg P(i)$ .}
\vspace{1ex}
\resizebox{.7\columnwidth}{!}{
\begin{tabular}{c|c|p{3.0cm}} 
  $g(i,o)$    & $Q(i,o)$  & Correction Action\\ \hline
\multirow{2}{*}{$\true$}
              & $\true$     & skip \\ \cline{2-3}
              & $\false$    & $Q \dn (i,o) \lor $ skip \\ \hline
\multirow{2}{*}{$\false$}
              & $\true$     & $P \dn (i) \lor S \dn (i,o)$ \\ \cline{2-3}
              & $\false$    & skip \\  \hline
\end{tabular}
}
\normalsize
\label{t:notpsqchecks}
\end{table}

We use the negative inputs to check whether the precondition 
should be weakened, and the postcondition
and the implementation should be strengthened. 

In the following listing of \cci{search} the precondition
guarantees the passed array \chci{a} to be sorted. 
This gives a chance to write a binary search
with a ${\cal O}(\log n)$ running time instead of the
${\cal O}(n)$ running time of \cci{linearSearch}.
However, our implementation is the same 
\cci{linearSearch} implementation.

\begin{tabular}{p{8cm}}
\begin{Verbatim}[fontsize=\relsize{-2},numbersep=4pt,numbers=left,frame=topline,framesep=4mm,
    label=\fbox{ Linear search for a sorted array},
commandchars=\\\{\}, codes={\catcode`$=3\catcode`_=8}]
int search(int [] a, int l, int r, int e) \{
  @pre srch \{ (0 $\leq$ l $\leq$ r < a.size)
     $\forall$ (int i:[1 .. a.size-1]) (a[i-1] $\le$ a[i]) \}
  int i = l;
  while ( i <= r ) \{
    if (a[i] == e)
      return i;
    i++; \}
  return -1;

   @post srch \{
      (0 $\leq$ l $\leq$ r < a.size)
      ( (rv $\not=$ -1) $\imp$ l $\leq$ rv $\leq$ r $\land$ a[rv] = e)
      ( (rv = -1) $\imp$ $\forall$ int k:[l .. r] (e $\not=$ a[k])) \}

  @behavior srch \{
    bad \{ input=\{a=\{1,3,5,4,2\}, l=0, r=4, e=2\} 
      output=\{rv=4\} \} $\ldots$ \} \}
\end{Verbatim}
\\
\end{tabular}

The \cci{bad} behavior on Line 17 passes an array \chci{a} that 
is not sorted and thus fails Line 3 in the precondition.
Also, the output passes the postcondition, however, the user deems the 
output as \bad since he expects a binary search behavior.
Table~\ref{t:notpsqchecks} suggests the corrections for the inputs
where $P(i)$ does not hold. 
In this case, the correction is either to weaken $P(i)$ to accept 
the unsorted array, or to modify the implementation.
We choose the latter since we
have not used the interesting precondition characteristics and
over implemented sacrificing a logarithmic efficiency gain 
in this case.

\begin{tabular}{p{8cm}}
\begin{Verbatim}[fontsize=\relsize{-2},numbersep=4pt,numbers=left,
    firstnumber=4,
    frame=topline,framesep=4mm,label=\fbox{ Binary search },
commandchars=\\\{\}, codes={\catcode`$=3\catcode`_=8}]
  $\ldots$
  if (r < l ) return -1;
  int mid = l + ( r  - l ) / 2;
  if ( a[mid] == e)
      return mid;
  if ( a[mid] > e )
      return search ( a,  e, l, mid - 1);
  return search (a, e, mid + 1, r); $\ldots$
\end{Verbatim}
\\
\end{tabular}

\section{The SpecCheck tool}
\label{s:tool}
\begin{figure}[t!]
\center{
\fbox{\parbox{4.7in}{
\begin{tabbing}
aai\=aaaai\=aai\=fffffffffffffffffffffffifffff\=\kill
{\bf SpecCheck($\{P\}S\{Q\}\{B\}$)}\\
\vspace{-.5em}
$\mathit{for~each~behavior}~b \in B$ \\
  \> $i$     \> $:=$ \> $b_{\mbox{input}}$; \> //~set input variables \\
  \>$P(i)$   \> $:=$\> $\mathit{traverse} (\mbox{pre})$; \> //~evaluate $P(i)$  \\
  \>$\mathit{apply}~\{P\}~\mathit{rules}$  \\
  \> $\mathit{if}~S~\mathit{is~empty}$  \\
aai\=aai\=aaaaaai\=aai\=fffffffffffffffffiffff\=\kill
  \> \>$o$     \> $:=$ \> $b_{\mbox{output}}$; \> //~set output variables \\
  \> \>$Q(i,o)$\> $:=$ \> $\mathit{traverse}~(\mbox{post})$; \> ~~//~evaluate $Q(i,o)$ \\
  \> \>$\mathit{apply}~\{P\}-\{Q\}\{B\}~\mathit{rules}$  \\
  \>$\mathit{else}$ \\
  \>  \> $o$ \> $:=$ \> $\mathit{traverse}(S)$; \> //~evaluate $S$, compute $o$ \\
  \> \>$Q(i,o)$\> $:=$ \> $\mathit{traverse} (\mbox{post})$; \\
  \> \>$\mathit{compute}~g(i,o)$; \\
  \> \>$\mathit{apply}~\{P\}S\{Q\}\{B\}~\mathit{rules}$
\end{tabbing}
\vspace{-1em}
} }
\caption{\bf Algorithm {\bf SpecCheck($\{P\}S\{Q\}\{B\}$) }}
\vspace{-.5em}
\label{a:speccheck}
}
\end{figure}

We built {\em SpecCheck} to evaluate our method. 
SpecCheck uses a simple ANTLR~\cite{Parr94antlr}
based front end to parse
implementations and specification triples ($\{P\}S\{Q\}\{B\}$)  
into a directed acyclic graph (DAG) representation where
nodes are programming elements, and edge weights describe 
operand relationships and statement order.

The algorithm of Figure~\ref{a:speccheck} illustrates how
SpecCheck computes the adequacy results for the properties.
First, it picks a behavior $b\in B$, and assigns the corresponding
input variables $i$ in the DAG 
to the input values 
$b_{\mbox{input}}$ of the behavior. 
Then it evaluates $P(i)$ with a recursive traversal of the DAG 
applying the semantics of every node in the DAG in the usual manner.
SpecCheck applies the correction rules to $P$
in accordance with Table~\ref{t:pchecks}, and waits for the user action to 
proceed.
The user may modify $P$, $Q$, and $S$, and may also append 
new behaviors to $B$. 
SpecCheck recomputes the DAG after the user intervention and
continues.
If the implementation $S$ is empty, SpecCheck assigns the 
corresponding output variables $o$ 
to the output values 
$b_{\mbox{output}}$ of the behavior,
and evaluates $Q(i,o)$ similarly
to $P(i)$ and the applies the $\{P\}-\{Q\}\{B\}$ rules from 
Table~\ref{t:qchecks} and waits for the user action to proceed. 
Again, SpecCheck recomputes the DAG after any intervention from the user.
If the implementation $S$ is not empty, SpecCheck computes $o$
by executing the implementation $S$, evaluates $Q(i,o)$,
and then computes $g(i,o)$ by either querying the user, or 
checking whether the values of $o$ match those of an existing good
behavior with the same $i$. 
Then SpecCheck applies the rules from Tables~\ref{t:psqchecks} 
and~\ref{t:notpsqchecks}.

The precondition and postcondition accept 
a specification name and an associated
predicate expression. 
No imperative constructs that modify variable values 
are allowed within the pre and post 
condition predicate expressions. 
SpecCheck supports a set of built-in predicates and
computable properties of programming constructs such as 
\cci{scasize}, \cci{scalpha}, \cci{scnum}, and \cci{scblank} 
that compute the array size, whether a scalar represents an ASCII 
alphabetical character, an ASCII numerical character, and an 
ASCII space, respectively.   

SpecCheck supports arithmetic and Boolean expressions, 
scalar and array data types, assignments, loops, and 
procedure calls including recursion.
It is a simple matter to extend SpecCheck to support 
other constructs; currently we support them through syntactic 
sugar. 
For example, we represent an array of C++ objects of type
\cci{ComplexNumber} with two scalar data fields as two arrays of
scalar numbers.

\section{The \cci{justify} example}
\label{s:justify}
To evaluate SpecCheck,  we used it to develop a more sophisticated 
program than \cci{linearSearch} with its specifications from scratch. 
We considered the \cci{justify} program that takes as input a single
paragraph of left-justified English text, 
and transforms it into a fully justified (both left and right justified) 
version by modifying the whitespaces within the paragraph.

Informally, the fully justified paragraph must satisfy the following requirements.
\bn
\i The non-whitespace text is not changed.

\i The output has a uniform line length of $W$ characters, for all
lines except possibly the last. This include spaces but not
the newline character.

\i The number of spaces between words on the same line are as 
uniform as possible.

\i A line should not contain more ``extra'' blanks than the length of
the first word of the following line.
\en

We started from the above informal requirements 
and built a number of behaviors that 
intuitively satisfy and test against them.
We quickly recognized that we need to define concepts such as
{\em blanks}, {\em words}, {\em lines}, and {\em paragraphs} formally.
We introduced a Boolean valued function for each concept
and listed behaviors that satisfy and test against 
each of the concepts.
Then we proceeded to build the pre and post conditions for each
function. 

We show below the code listing for the function \cci{sameWords}
that takes as input two strings 
\cci{p1} and \cci{p2} with two offsets \cci{l1} and \cci{l2},
and claims that the two paragraphs starting at the offsets 
have the same non-whitespace contents.
The two offsets serve well to define the function recursively. 
The full \cci{justify} example is available 
online$^{\ref{fn:online}}$.

\begin{tabular}{p{8cm}}
  \begin{Verbatim}[fontsize=\relsize{-2},numbersep=4pt,numbers=left,
    frame=topline,framesep=4mm,
    label=\fbox{ sameWords Boolean function },
commandchars=\\\{\}, codes={\catcode`$=3\catcode`_=8}]
boolean sameWords(int[] p1, int[] p2, int l1, int l2) \{
  @pre swspec \{ 
    subIsPara(p1, l1); 
    subIsPara(p2, l2); \}
\end{Verbatim}
\\
\end{tabular}

The \cci{swspec} precondition guarantees that the
text that starts at \cci{l1} and \cci{l2} 
in \cci{p1} and \cci{p2}, respectively, is a paragraph.
We define the \cci{subIsPara(int [] str, int l)} function recursively 
with its specification and behaviors.
The base case of \cci{subIsPara} checks whether the 
\cci{str[l:str.size-1]}
, the substring of 
\chci{str} that starts at index \chci{l} and ends at index 
\chci{str.size-1} inclusive, 
is a line, 
and the inductive case checks
whether it is a line followed by a paragraph.

To claim that \cci{p1} and \cci{p2} have the same contents,
we invoke the function \cci{sameWords(p1,p2,0,0)} as in the first
\cci{good} behavior below.

\begin{tabular}{p{8cm}}
  \begin{Verbatim}[fontsize=\relsize{-2},numbersep=4pt,numbers=left,firstnumber=38,
commandchars=\\\{\}, codes={\catcode`$=3\catcode`_=8}]
    $\ldots$
@behavior swspec \{
  good \{input=\{p1="aaN",p2="aaN",l1=0,l2=0\};
    output=\{rv=true\}\};
  bad  \{input=\{p1="aN",p2="aaN",l1=0,l2=0\};
    output=\{rv=true\}\};
  good \{input=\{p1="aa aaaNaaa  aaN",
           p2="aa  aaaNaaa aaN",l1=0,l2=0\};
        output=\{rv=true\}\};
  good \{input=\{p1="aaa  aa aaaNaaa  aaN",
           p2="aaa aa  aaaNaaa aaN",l1=5,l2=4\};
        output=\{rv=true\}\};
  good \{input=\{p1="aaa  aa aaaNaaa  aaN",
           p2="aaa aa  aaaNaaa aaN",l1=5,l2=4\};
        output=\{rv=true\}\};
  good \{input=\{p1="aaa  aaaa aaaNaaa  aaN",
           p2="aaa aa  aaaNaaa aaN",l1=5,l2=4\};
       output=\{rv=false\}\};
  bad  \{input=\{p1="a aaNaaN",
           p2="a  aaNaaaN",l1=0,l2=0\};
        output=\{rv=true\}\}; \}
\end{Verbatim}
\\
\end{tabular}

The \chci{a} characters in the input strings designate the 
alphanumeric and punctuation characters, and the \chci{N} 
character designates the newline character. 
We used the five good and two bad behaviors presented above
in a fashion similar to \cci{linearSearch} 
to come up with the following postcondition of
\cci{sameWords}. 

\begin{tabular}{p{8cm}}
  \begin{Verbatim}[fontsize=\relsize{-2},numbersep=4pt,numbers=left,firstnumber=22,
commandchars=\\\{\}, codes={\catcode`$=3\catcode`_=8}]
    $\ldots$
@post swspec \{
  subIsPara(p1, l1);
  subIsPara(p2, l2);
  ( p1=p2 &&                 // p1 = p2 = word-newline
      subIsWord(p1,0,p1.size -2) && 
      p1[p1.size-1]==newline) 
||
  exists (int w1:[l1 .. p1.size-1],  //there exists 
      int w2:[l2 .. p2.size-1],) \{   //a partition 
  exists (int k1:[w1+1 .. p1.size-1],//of p1, p2 
      int k2:[w2+1 .. p2.size-1]) \{  //such that
    p1[l1:w1]=p2[l2:w2] &&           //p1=word-q1, 
    headTailOfSub(p1,l1,w1,k1) &&    //p2=word-q2,
    headTailOfSub(p2,l2,w2,k2) &&    //sameWords(q1,q2)
    sameWords(p1,p2,(k1+1),(k2+1)) \} \}; \}
\end{Verbatim}
\\
\end{tabular}

The base case of \cci{sameWords} claims that
\cci{p1} and \cci{p2} are equal and consist of a word
followed by a newline. It uses the 
function \cci{subIsWord(int[] str, int l, int r)} 
that returns $\true$ when \cci{a[l:r]}
consists of 
alphanumeric characters only. 

\begin{figure*}
\centering
\resizebox{1.0\columnwidth}{!}{
\includegraphics{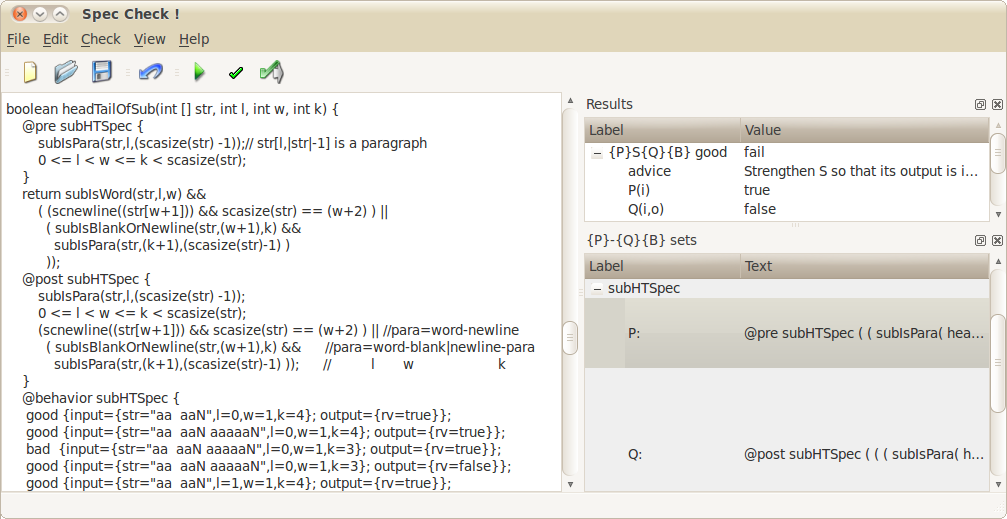} }
\caption{Snapshot of the Spec~Check~! tool with the \cci{justify} example.}
\label{f:checkspec}
\vspace{-1em}
\end{figure*}

The inductive case partitions the substrings 
starting at \cci{l1} and \cci{l2} in \cci{p1} and \cci{p2}, 
respectively, 
into two $\langle hd,blank,tail\rangle$ tuples where
the heads are equal, \cci{p1[l1:w1] = p2[l2:w2]}, 
the blanks (\cci{p1[w1+1:k1]} and \cci{p2[w2+1:k2]})
may be of different lengths, 
and the tails have the same non-whitespace
content. 
The function \cci{headTailOfSub} returns $\true$ 
when its parameters define such a tuple.

The behaviors and the correction actions were instrumental 
to come up with the final version of the \cci{sameWords}
specification.
In particular, the behaviors with varying blank spaces, 
helped fix mistakes with the indices. 
The behaviors with single words on a line helped us to refine
our recursive definitions of a line and a paragraph (\cci{subIsLine} and \cci{subIsPara})
to use the single word as the base case rather than the empty string $\epsilon$.
This is good for two reasons:
\be
\i Allowing empty lines and paragraphs make some function definitions return
more than one value, \ie they would define a relation and not a function. 
This is awkward and requires repeatedly adding conjuncts
such as $w \ne \epsilon$ and $\l \ne \epsilon$ and results in longer,
more verbose and prone to mistakes definitions. 
Behaviors directed us to abide by an important design principle 
since it is more concise to state the non-emptiness requirement 
in one place: the definitions of word and line.

\i Empty lines and paragraphs do not correspond to our intuition of what a single
paragraph looks like.
We believe that the intuitiveness of the definition came as a result 
of the intuitiveness of the provided behaviors.
\ee

The use of substring indices in the function definitions 
helped eliminate some
quantifiers from our predicates which in turns helps with the accuracy checks 
of the specification.

The implementation of the \cci{sameWords} functions was exactly similar to the 
specification since it serves as a predicate function. 
This was not the case for the imperative functions such as \cci{int head(int [])}, 
and \cci{int tail(int [])} that compute the indices of the head and tail of
a given paragraph.

We briefly describe the rest of the \cci{justify} specifications 
and implementation below. 
The precondition of the \cci{justify(int [] str, int $W$)} function accepts
single paragraphs or empty strings \cci{str = $\epsilon$} $\lor$ \cci{subIsPara (str,0)}.
A behavior with a word on a line longer than $W$ helped us modify the precondition to specify
that all words must be of length $\le W$. 
This happened because we used $W$ to be small in order to be able to practically interact with 
the tool. 
Another behavior with a word with length $W-2$ followed by a word of length $>2$ 
lead us to refine the precondition and only allow words with length $< W/2$.

In addition to the \cci{sameWords} function, we used several other 
functions to define the postcondition of \cci{justify}.
For example, the \cci{minSep} and \cci{maxSep} functions compute the minimum 
and maximum inter-word separation of a line and the \cci{justLength} Boolean function 
restricts their difference to be $\le 1$.
This expresses the space line uniformity.

The \cci{goodSep} function recursively computes the total excess separation and the 
postcondition makes sure it is smaller than the length of the head of the next line. 
It uses \cci{numWords} and \cci{totalSep} to compute the number of words and the 
total number of blank characters on a line respectively. 
The function \cci{goodSep} defines the excess separation as the 
difference as there must be at least $n-1$ separators between $n$ words on a line. 

Figure~\ref{f:checkspec} shows a snapshot of the SpecCheck
graphical user interface (GUI)
with results for the \cci{justify} example. 
We provide the full implementation and the specification of \cci{justify}
along with SpecCheck online$^{\ref{fn:online}}$.

\section{Related work }
\label{s:related}




\label{sec:related}

Methods for checking specifications against behaviors have been
presented in the literature \cite{HWT03,HST98,HKLAB98,KKK95,CABetal98}. These check a
specification against behaviors by first writing the specification and
then attempt to verify if the specification is
accurate using a method such as animation, execution, model-checking, etc.
In contrast, we go in the other direction: we write the
specification from the behaviors, so that the specification is
accurate by construction. In \cite{PP03}, a method for writing
trace-based specifications is presented. The behaviors are sequences
of atomic events. The technique is suitable for specifying reactive modules as
``black boxes'' that interact with an environment via events. Our
method, in contrast, views programs as white boxes, and our
specification are pre/postcondition pairs over the program state. Our
method is intended for transformation programs that perform nontrivial
computations on data, rather than reactive modules where control is
the major issue. 
A method of writing temporal-logic based specifications using event
traces (``scenarios'') is presented in \cite{LW98}. This also applies
to reactive systems and stresses control rather than data.
in \cite{F89}, a method for refining an initially simple specification
using ``elaborations'' is presented. The elaborations are presented
informally. A method of checking software cost reduction (SCR) specifications for consistency 
is presented in \cite{HJL96}. 

In none of the above works is there any analogue to our correction
actions, and in particular to the construction of a correction action
as a Boolean formula, which can then be carried out in several
different ways (due to or and xor operators), depending on the
developers intuition.
The SPECIFIER \cite{MH91} tool constructs formal specifications of
data types and programs from informal descriptions, but uses schema,
analogy, and difference-based reasoning, rather than our approach which
is directly based on input-output behaviors.  The Larch \cite{GGH90}
approach enables the verification of claims about specifications,
using the Larch prover, which improves the confidence in the
specification's accuracy.
In \cite{MS03}, a method for testing preconditions, postconditions,
and state invariants, using mutation analysis, is presented. 
It would be interesting to compare this method with our approach of
checking a specification against a finite-domain version of the same specification.

\section{Conclusion}
\label{s:conclusion}
We presented a method to manually construct a specification from 
use cases and two algorithms for checking the accuracy of the 
resulting check.
We illustrated our method with two examples; \cci{linearSearch} and 
\cci{justify}. 
Future theoretical development consists of extending the class of 
formulae that can be checked formally by the reduction algorithm.
Currently this class is expressive enough to specify properties 
such as sortedness, injectivity, and selectedness.
Future applications consist of using the SpecCheck tool in 
undergraduate courses to introduce students to formalism and to 
help them write formal specifications correctly.

\clearpage
\raggedbottom

\bibliographystyle{IEEEtran}

\bibliography{specCheck}

\begin{thebibliography}{10}
\providecommand{\url}[1]{#1}
\csname url@samestyle\endcsname
\providecommand{\newblock}{\relax}
\providecommand{\bibinfo}[2]{#2}
\providecommand{\BIBentrySTDinterwordspacing}{\spaceskip=0pt\relax}
\providecommand{\BIBentryALTinterwordstretchfactor}{4}
\providecommand{\BIBentryALTinterwordspacing}{\spaceskip=\fontdimen2\font plus
\BIBentryALTinterwordstretchfactor\fontdimen3\font minus
  \fontdimen4\font\relax}
\providecommand{\BIBforeignlanguage}[2]{{%
\expandafter\ifx\csname l@#1\endcsname\relax
\typeout{** WARNING: IEEEtran.bst: No hyphenation pattern has been}%
\typeout{** loaded for the language `#1'. Using the pattern for}%
\typeout{** the default language instead.}%
\else
\language=\csname l@#1\endcsname
\fi
#2}}
\providecommand{\BIBdecl}{\relax}
\BIBdecl

\bibitem{Dij76}
E.~Dijkstra, \emph{A Discipline of Programming}.\hskip 1em plus 0.5em minus
  0.4em\relax Englewood Cliffs, New Jersey: Prentice-Hall Inc., 1976.

\bibitem{Ho69}
C.~Hoare, ``An axiomatic basis for computer programming,'' \emph{Commun. ACM},
  vol.~12, no.~10, pp. 576--580, 583, 1969.

\bibitem{Gr81}
D.~Gries, \emph{The Science of Programming}.\hskip 1em plus 0.5em minus
  0.4em\relax New York: Springer Verlag, 1981.

\bibitem{HWT03}
M.~Heimdahl, M.~Whalen, and J.~Thompson, ``Nimbus: a tool for specification
  centered development,'' in \emph{Requirements Engineering Conference, 2003.
  Proceedings. 11th IEEE International}, Sept. 2003, p. 349.

\bibitem{HST98}
D.~Hazel, P.~Strooper, and O.~Traynor, ``Requirements engineering and
  verification using specification animation,'' in \emph{Automated Software
  Engineering. Proceedings. 13th IEEE International Conference on}, oct 1998,
  pp. 302 --305.

\bibitem{HKLAB98}
C.~Heitmeyer, J.~Kirby, J., B.~Labaw, M.~Archer, and R.~Bharadwaj, ``Using
  abstraction and model checking to detect safety violations in requirements
  specifications,'' \emph{Software Engineering, IEEE Transactions on}, vol.~24,
  no.~11, pp. 927 --948, nov 1998.

\bibitem{KKK95}
E.~M. Kim, S.~Kusumoto, and T.~Kikuno, ``An approach to safety and correctness
  verification of software design specification,'' in \emph{Software
  Reliability Engineering}, oct 1995, pp. 78 --83.

\bibitem{CABetal98}
W.~Chan, R.~Anderson, P.~Beame, S.~Burns, F.~Modugno, D.~Notkin, and J.~Reese,
  ``Model checking large software specifications,'' \emph{Software Engineering,
  IEEE Transactions on}, vol.~24, no.~7, pp. 498 --520, jul 1998.

\bibitem{F89}
M.~Feather, ``Constructing specifications by combining parallel elaborations,''
  \emph{Software Engineering, IEEE Transactions on}, vol.~15, no.~2, pp. 198
  --208, feb 1989.

\bibitem{HJL96}
C.~L. Heitmeyer, R.~D. Jeffords, and B.~G. Labaw, ``Automated consistency
  checking of requirements specifications,'' \emph{ACM Trans. Softw. Eng.
  Methodol.}, vol.~5, pp. 231--261, July 1996.

\bibitem{MH91}
K.~Miriyala and M.~Harandi, ``Automatic derivation of formal software
  specifications from informal descriptions,'' \emph{Software Engineering, IEEE
  Transactions on}, vol.~17, no.~10, pp. 1126 --1142, oct 1991.

\bibitem{GGH90}
S.~Garland, J.~Guttag, and J.~Horning, ``Debugging larch shared language
  specifications,'' \emph{Software Engineering, IEEE Transactions on}, vol.~16,
  no.~9, pp. 1044 --1057, sep 1990.

\bibitem{MS03}
T.~Miller and P.~Strooper, ``A framework and tool support for the systematic
  testing of model-based specifications,'' \emph{ACM Transactions on Software
  Engineering Methodologies}, vol.~12, pp. 409--439, October 2003.

\bibitem{Parr94antlr}
T.~J. Parr and R.~W. Quong, ``Antlr: A predicated-ll(k) parser generator,''
  \emph{Software Practice and Experience}, vol.~25, pp. 789--810, 1994.

\bibitem{PP03}
S.~Prowell and J.~Poore, ``Foundations of sequence-based software
  specification,'' \emph{Software Engineering, IEEE Transactions on}, vol.~29,
  no.~5, pp. 417 -- 429, may 2003.

\bibitem{LW98}
A.~van Lamsweerde and L.~Willemet, ``Inferring declarative requirements
  specifications from operational scenarios,'' \emph{Software Engineering, IEEE
  Transactions on}, vol.~24, no.~12, pp. 1089 --1114, dec 1998.

\end{thebibliography}

\end{document}